\documentclass[twocolumn]{aastex631}
\usepackage{amsmath}
\usepackage{natbib}

\usepackage{float}
\usepackage{multirow}
\usepackage{array}
\usepackage{booktabs}

\begin{document}

\title{Non-detection of FAST and Parkes follow-up observation for 27 Parkes discovered FRBs}

\correspondingauthor{Xuan Yang, Songbo Zhang, Xuefeng Wu}
\email{yangxuan@pmo.ac.cn, sbzhang@pmo.ac.cn, xfwu@pmo.ac.cn}

\author{Xuan Yang}
\affiliation{Purple Mountain Observatory, Chinese Academy of Sciences, Nanjing 210023, China}
\affiliation{School of Astronomy and Space Sciences, University of Science and Technology of China, Hefei 230026, China}

\author{Songbo Zhang}
\affiliation{Purple Mountain Observatory, Chinese Academy of Sciences, Nanjing 210023, China}
\affiliation{CSIRO Space and Astronomy, Australia Telescope National Facility, PO Box 76, Epping, NSW 1710, Australia}

\author{Xuefeng Wu}
\affiliation{Purple Mountain Observatory, Chinese Academy of Sciences, Nanjing 210023, China}
\affiliation{School of Astronomy and Space Sciences, University of Science and Technology of China, Hefei 230026, China}


\begin{abstract} %up to 250 words
To investigate whether apparently non-repeating Fast Radio Bursts (FRBs) are truly one-off transients, we conducted systematic follow-up observations of 27 out of 81 non-repeating FRBs identified in the Parkes Transient Database. Using 59.0 hours of data from the Parkes Ultra-Wideband Low (UWL) receiver and 6.3 hours from the Five-hundred-meter Aperture Spherical Telescope (FAST) 19-beam receiver, we searched for repeated bursts from these sources. No additional bursts were detected from any of the 27 FRBs. Combining these non-detections with prior archival observations, we derived stringent upper limits on their repetition rates above 1 Jy under two statistical models: Poisson process constraints range from $\sim10^{-3.5}$ to $10^{-1.9}\,\mathrm{h^{-1}}$, while Weibull process constraints range from $\sim10^{-3.4}$ to $10^{-1.5}\,\mathrm{h^{-1}}$. These limits are approximately an order of magnitude stricter than those reported in previous studies. By applying consistent observational setups and analytical methodologies across all sources, the derived rate limits converge to a narrow, well-defined range. This suggests that these FRBs form a relatively homogeneous population with extremely low intrinsic activity rates.
%In this work, we present follow-up observations of 27 fast radio bursts (FRBs) detected from archival data of the Parkes radio telescope between 1997 and 2001. These follow-up observations were conducted with the Parkes radio telescope Ultra-Wideband Low receiver and the Five-hundred-meter Aperture Spherical Telescope (FAST) 19-beam receiver. No additional bursts from these sources were detected. Based on these nondetections, we provide constraints on the repetition rate for 27 sources. We constrain the repetition rates of these sources using two statistical models of FRB occurrence. Combining our non-detections with prior observations, we derive upper limits on the repetition rates above 1 Jy of the sources from $\sim 10^{-3.5}$ to $10^{-1.9} h^{-1}$ under a Poisson process, and $\sim 10^{-3.4}$ to $10^{-1.5} h^{-1}$ under a Weibull process.
\end{abstract}

%\keywords{Radio bursts (1339), Radio transient sources (2008), Radio pulsars (1353), Polarimetry (1278)}

\section{Introduction}

Fast Radio Bursts (FRBs) are bright radio pulses with a short duration of typical milliseconds. 
Since the first FRB was discovered in Parkes archival observation~\citep{Lorimer2007qn}, more than 800 FRBs have been detected by telescopes worldwide~\footnote{The FRB online Catalogue is available from \url{www.wis-tns.org}}. %~\citep{Blinkverse}
Their dispersion measures (DMs) typically exceed the contribution expected from the Milky Way~\citep{ne2001,Yao2017}, and they are considered to be extragalactic. More than 110 FRB sources have been localized to their extragalactic host galaxies to date~\citep[e.g.][]{Chatterjee17,askap_frb190711,host_frb20240209A}. However, a notable exception is FRB~200428, which was associated with a Galactic magnetar and has a lower luminosity compared to most extragalactic FRBs~\citep{STARE2,frb200428}.

%In addition to their origins, 
Another key observational property of FRBs is their repeatability. While the majority have only been detected once,
nearly 70 sources have exhibited repeated bursts~\citep{121102_2,CHIME19_8r,Blinkverse,MeerKATrepeat}. FRBs are therefore often categorized into two groups based on their apparent repeatability.
%leading to the division of FRBs into two classes based on their apparent repeatability. 
%As the population of FRBs has rapidly increased and their properties have been studied in more detail, a better understanding of their classification has emerged.
%For example, \cite{volume_rate} concluded that most FRB events must originate from repeaters based on the evidence of their volumetric occurrence rate. Additionally,  \cite{energy_distribution} found that the Canadian Hydrogen Intensity Mapping Experiment (CHIME) repeater sample is consistent with the energy dependence of the apparently non-repeating Australian Square Kilometre Array Pathfinder~(ASKAP) sample, suggesting that they may come from the same population.

%The data from the Parkes radio telescope, which played a important role in the early discovery and characterization of FRBs, are essential to the discussion of FRB repeatability~\citep[e.g.][]{parkes4frb,parkes4frb2}. 
The Parkes radio telescope played an important role in the early discovery and characterization of FRBs, and its data remain essential for studies of FRB repeatability~\citep[e.g.][]{parkes4frb,parkes4frb2}. 
A significant fraction of Parkes-detected FRBs have not been observed to repeat despite various follow-up observations. 
In particular, among the 30 bright FRBs detected by Parkes with a signal-to-noise ratio (S/N) above 10, only FRB~20180301A has been confirmed as a repeater, through follow-up observations using the FAST and Parkes telescopes~\citep{frb180301a1,frb20180301a2,songborrat}. 
It therefore remains unclear whether the apparently non-repeating sources represent genuinely one-off events or instead belong to a population with extremely low burst rates. Some studies suggest that some of these sources may be weakly active repeaters~\citep{good2023}.
%Whether these apparently non-repeating sources are genuinely one-off events or simply exhibit very low repetition rates remains an open question. Some studies suggest that a subset of apparently non-repeating FRBs may in fact belong to a low-activity repeating population~\citep{good2023}.

%We analyzed and concluded their extraterrestrial origin using both multibeam detection status and statistical perspective~\citep[see more details in][]{81candi}.
Besides the bright FRBs, Parkes has also detected a population of relatively faint events~\cite{81candi}.
In \citet{Zhang_2020}, we reprocessed all observations obtained during the first four years of operation (1997$-$2001) with the Parkes Multibeam Receiver. A total of 568,736,756 pulses with a S/N larger than seven were recorded in a single pulse database, known as the Parkes Transient Database I~(PTD-I). 
In our subsequent work \cite{81candi}, we applied a machine learning classification to all events in PTD-I and identified 81 apparently one-off FRBs with S/N values ranging from 7.1 to 8.2. 
Based on their detection in a single beam of the multibeam system, together with their DMs exceeding the expected Galactic contribution, we concluded that these events are of extragalactic origin~\citep[see more details in][]{81candi}.

The Five-hundred-meter Aperture Spherical radio Telescope (FAST), with its exceptional sensitivity~\citep{fast_parameter}, provides a unique opportunity to detect fainter bursts and better constrain the repetition behavior of FRBs. Although several sensitive follow-up campaigns with FAST and other sensitive instruments have already placed stringent upper limits on the repetition rates of part of non-repeating FRB candidates from the CHIME/FRB catalog~\citep{fast36repeat}, a systematic study specifically targeting the Parkes-discovered sample is still lacking. 
%The Parkes-discovered sample is worth studying because it represents a homogeneously identified group using the same instrument and pipeline. It is ideal for testing FRB population characteristics, free from the biases introduced by comparing results from different telescopes and surveys. 
%

Parkes accumulated the first substantial sample before the CHIME era~\citep{chimecatalog}. During this pioneering period, the Arecibo telescope discovered FRB~121102 in 2014~\citep{121102_1}, which was later confirmed as the first repeating FRB~\citep{121102_2},  while the most Parkes discovered FRBs remained apparently one-off events~\citep{songborrat}. This striking contrast raises important questions about whether this reflects intrinsic differences in source populations, observational selection effects, or simply insufficient follow-up sensitivity.
Investigating the repetition properties of these historically accumulated, apparently one-off Parkes FRBs with more sensitive observations will provide valuable insights into this temporally distinct sample.

%sub-groups 
%With the rapid growth of the FRB samples, possible sub-populations are gradually arising among the repeating or apparently non-repeating FRBs. Most of the repeaters are highly linearly polarized~\citep[][]{121102_polar,linear_polarisation}, but FRB180301 shows a diversity of polarimetric roperties~\citep[][]{luoruifrb}. Only two out of the nearly 30 repeaters show a periodicity~\citep[][]{121102_period,180916_period}. The repetition rates of some repeaters are reported much lower than those notable repeating FRBs like FRB 20121102A and FRB 20201124A~\citep[][]{good2023,fast36repeat}. \cite{frbtype} apply Kolmogorov-Smirnov (KS) test in CHIME/FRB catalogue, and introduce four new sub-types of repeaters and apparently non-repeaters with respect to the cosmic star formation history (SFH). However, the final confirmation of any sub-populations for the repeating or apparently non-repeating FRBs still needs more completed samples and higher sensitivity observations.

%In this work, we report follow-up observations of 27 apparently one-off FRBs with the FAST and the Parkes telescope. We report the upper limits on their repetition rates, derived from the Poisson process and Weibull process.

In this work, we present follow-up observations of 27 FRBs from the Parkes Transient Database, using both FAST and the Parkes latest ultrawide-bandwidth low-frequency (UWL) receiver to search for repeated bursts from these sources.
%using both FAST and the Parkes telescope itself. 
%and to place constraints on their repetition rates. 
%Based on the non-detections, we derive upper limits on the repetition rates under both Poisson and Weibull models.
In Section~\ref{sec:data}, we provide a description of the data and searching pipeline used in this study. 
%The methods of our rate analysis are presented in Section~\ref{sec:methods}, while 
Section~\ref{sec:discuss} includes the results and discussion of our observations. Section~\ref{sec:Conclusions} concludes this work.

\section{Observation and Data Reduction} \label{sec:data}

The 81 FRBs identified in PTD-I were obtained from Parkes Multibeam observations~\citep{multibeam}, which had a central frequency of 1374~MHz, a bandwidth of 288~MHz, and 96 channels~\footnote{Detailed parameters for the 81 FRB sources are available at~\url{https://astroyx.github.io/}.}. 
We aimed to monitor as many of these 81 FRBs as possible; however, due to limitations in available observing time, we prioritized targets based on their S/N and observing cadence.
Consequently, 27 FRBs were selected for follow-up observations using FAST and Parkes telescopes. The total integration time and observing parameters for each FRB are summarized in Table~\ref{table:obsproperties}.

For Parkes observation, we used the UWL receiver~\citep{uwlreceiver}. Its lowest band ($\sim$768\,MHz) has a half maximum power (FWHM) of $\sim0.4^\circ$, nearly twice that of the previous multibeam receiver ($\sim0.23^\circ$) where our targeted sources were initially detected.  
The UWL receiver's bandwidth of 3328\,MHz also represents a significant improvement over the previous multibeam observation’s 288\,MHz, providing a better chance of detecting repeated signals even if their central frequency shifts. The feasibility of this strategy is demonstrated by our new detection of repeating pulses from an apparent one-off Rotating Radio Transient (RRAT)~\citep{songborrat}.
A total of 59.0 hours of Parkes UWL observations were recorded, targeting 21 FRBs between 2020-12-26 and 2025-01-23. The complete observation log is listed in Table~\ref{table:obslog}.
Observations were conducted with a central frequency of 2368 MHz, a bandwidth of 3328 MHz, 13312 frequency channels, a sample time of 64 $\rm \mu s$,  and 2-bit sampling.

For FAST, a total of 6.3 hours observations were conducted, targeting 6 FRBs on 2021-08-09, 2021-08-13, 2025-08-24, 2025-09-05. We used the FAST 19-beam receiver in Snapshot mode, which was developed to efficiently map the sky using a four-point gridding strategy~\citep{fast_parameter}. The snapshot sky coverage of $\sim 0.5^\circ$ in diameter ensured that the initial Parkes multibeam localization uncertainties were fully covered.
The central frequency, bandwidth, the number of channels , sample time and sampling bit for these observations were 1250 MHz, 500 MHz, 4096, and 196.608 $\rm \mu s$, 8-bit, respectively.

The data were processed using the single-pulse search software from \emph{\sc PRESTO}~\citep{2001PhDT123R, 2002AJ1241788R,2011asclsoft07017R}~\footnote{\url{http://www.cv.nrao.edu/~sransom/presto/}}. 
We searched the full-band FAST data, while the Parkes UWL data were divided into a series of sub-bands from 128 to 3328\,MHz following a tiered search strategy~\citep{frb20180301a2,frb0529}.
The raw data were first masked using the \emph{\sc rfifind} script in \emph{\sc PRESTO} to mitigate radio-frequency interference~(RFI). During the \emph{\sc prepdata} process, a DM step of 0.1 pc$\cdot$cm$^{-3}$ was adopted for dedispersion, and the searched DM range spanned from –20 to +20 pc$\cdot$cm$^{-3}$ relative to the known DM of each source, which are commonly used values in known FRB searching pipelines like~\cite{frbsearch}. 
All dedispersed data were then searched using the \emph{\sc single\_pulse\_search.py} script, with a S/N threshold set to 7 and boxcar widths of 1, 2, 3, 4, 6, 9, 14, 20, 30, 45, 70, 100, 150, 220, and 300 samples, which is consistent with \cite{Zhang_2020}. The searching parameters would result in a fluence detection limit of 0.03 and 0.006 Jy\,ms for Parkes and FAST respectively, assuming a burst with width close to the sampling time. The effectiveness and reliability of this pipeline have been demonstrated by the successful detection of over 1000 bursts from similar FAST and Parkes UWL observations of FRB~20220529~\citep{frb0529}.

\begin{table*}[ht]
  \begin{scriptsize}
  \caption{The observation properties of 27 FRBs, sources observed by FAST are marked with an asterisk ($^*$) in the source name column. The total PTD-I integration time is 50.0 hours. The follow up integration time is 59.0 hours for Parkes and 6.3 hours for FAST. The DM\_exc is the DM value exceeding the contribution of the Milky Way~\citep{Yao2017}. \label{table:obsproperties}}
  \renewcommand\arraystretch{1.0}
  \setlength{\tabcolsep}{2.0mm}  
  \begin{center}
  \begin{tabular}{@{\extracolsep{\fill}}ccccccccc}
  \hline  
\multicolumn{1}{c}{Source}  & PTD-I  &     Follow up     &  DM        &     DM\_exc    &     Initial  & Initial  &   R.A. & Dec.    \\
 \multicolumn{1}{c}{Name}   &Integration Time(h) & Integration  Time(h)   &  (pc$\cdot$cm$^{-3}$) &    (pc$\cdot$cm$^{-3}$)    &  S/N & Width(ms) & (J2000)  & (J2000) \\
\hline
FRB20010702A* & 0.22 & 2.00 & 182.3 & 109.0 & 7.8 & 5.4 & 17:50:37.42 & 11:54:00.55 \\ 
FRB20010703C* & 0.22 & 2.00 & 177.1 & 92.7 & 7.7 & 9.4 & 18:18:34.50 & 19:56:02.67 \\ 
FRB20010630E* & 0.37 & 0.58 & 187.1 & 110.2 & 7.5 & 3.4 & 18:15:50.77 & 21:33:42.30 \\ 
FRB20010630D* & 0.15 & 0.58 & 82.4 & 12.4 & 7.5 & 5.4 & 18:10:50.85 & 22:27:12.41 \\ 
FRB20010703B* & 0.22 & 0.58 & 226.7 & 159.7 & 7.3 & 10.1 & 17:58:05.30 & 18:26:49.72 \\ 
FRB20010621C* & 1.17 & 0.58 & 259.1 & 4.6 & 7.2 & 14.1 & 19:20:38.50 & 5:20:34.33 \\ 
FRB20010303A & 0.15 & 6.86 & 214.8 & 185.8 & 7.2 & 7.5 & 4:09:10.27 & -44:29:06.27 \\ 
FRB20001109A & 2.33 & 6.02 & 528.9 & 498.8 & 8.0 & 17.7 & 1:35:18.05 & -72:01:17.59 \\ 
FRB20010612A & 4.67 & 5.24 & 422.9 & 390.0 & 7.4 & 11.3 & 1:39:32.23 & -75:01:24.18 \\ 
FRB20000622C & 2.33 & 5.12 & 382.5 & 351.2 & 7.4 & 11.2 & 0:53:05.17 & -73:59:53.00 \\ 
FRB20001222A & 7.00 & 4.78 & 508.9 & 450.6 & 7.5 & 21.0 & 5:31:52.80 & -66:44:56.61 \\ 
FRB19990427A & 1.75 & 4.53 & 905.5 & 436.7 & 7.7 & 8.8 & 13:42:01.51 & -59:25:54.39 \\ 
FRB20000928A & 4.67 & 3.52 & 853.0 & 805.8 & 7.5 & 40.3 & 4:46:50.18 & -67:15:29.16 \\ 
FRB20001118A & 0.66 & 3.51 & 730.2 & 424.7 & 8.0 & 19.9 & 17:04:59.70 & -49:51:28.14 \\ 
FRB20001123A & 7.00 & 3.25 & 368.6 & 320.6 & 8.1 & 18.4 & 5:00:39.99 & -64:57:46.22 \\ 
FRB20010509A & 7.52 & 2.72 & 592.3 & 541.5 & 7.6 & 26.7 & 5:00:04.51 & -67:55:11.36 \\ 
FRB20010125C & 0.22 & 2.50 & 751.2 & 646.2 & 8.1 & 10.6 & 12:11:13.76 & -43:30:58.15 \\ 
FRB20010625A & 0.22 & 2.31 & 479.7 & 404.0 & 7.7 & 9.9 & 8:59:42.63 & -22:19:07.81 \\ 
FRB20010128A & 0.22 & 1.51 & 225.9 & 121.9 & 7.3 & 6.7 & 10:58:43.41 & -81:05:48.26 \\ 
FRB20000925A & 2.33 & 1.29 & 206.6 & 152.0 & 7.5 & 20.2 & 5:09:54.96 & -69:59:49.70 \\ 
FRB20010127A & 0.22 & 1.19 & 126.8 & 42.2 & 7.2 & 3.7 & 10:20:02.86 & -32:24:18.73 \\ 
FRB19980624A & 0.30 & 1.00 & 211.4 & 175.2 & 7.6 & 10.3 & 4:53:38.17 & -53:49:47.47 \\ 
FRB19990511A & 0.29 & 0.92 & 410.0 & 234.4 & 7.7 & 10.1 & 14:12:21.66 & -72:36:22.19 \\ 
FRB20010215A & 5.30 & 0.72 & 1131.4 & 1065.0 & 7.2 & 39.6 & 5:51:45.00 & -71:14:01.41 \\ 
FRB20010819A & 0.15 & 0.71 & 114.2 & 59.4 & 7.8 & 18.3 & 5:38:49.26 & -28:10:59.70 \\ 
FRB20010127B & 0.15 & 0.70 & 171.9 & 23.8 & 7.2 & 3.6 & 10:52:59.72 & -43:03:39.39 \\ 
FRB20010316A & 0.15 & 0.66 & 478.5 & 432.7 & 7.7 & 13.8 & 5:34:42.22 & -37:02:41.69 \\ 
\hline
\end{tabular}
\end{center}
\end{scriptsize}
\end{table*}

\section{Methods}\label{sec:methods}

\subsection{Sensitivity limit}
The flux density limit can be estimated using the radiometer equation~\citep{Manchester2001,dmsmearing}:
\begin{equation}
    S=\frac{\sigma \  S/N \ T_{\rm sys}}{G W_i}\sqrt{\frac{W_b}{ \Delta \nu  N_p}},
	\label{eq:slim}
\end{equation}
where $\sigma$ is a loss factor due to digitisation, which is $\sim 1$ for our observations and consistent with~\cite{good2023} and~\cite{fast36repeat}. $T_{\rm sys}$ is the system temperature: $\sim$21~K for the Parkes multibeam receiver, and $\sim$23~K for both the Parkes UWL receiver and the FAST 19-beam receiver. $G$ is the telescope antenna gain, with values of $\sim$0.735, 0.757, and 16~K/Jy for the Parkes multibeam, Parkes UWL, and FAST 19-beam receivers, respectively~\footnote{The $T_{\rm sys}$ and $G$ values for the Parkes multibeam, UWL and FAST 19-beam receivers are referenced from~\cite{Manchester2001,uwlreceiver} and~\cite{fast_parameter}, respectively.}. $\Delta \nu$ is the observing bandwidth as described in Section~\ref{sec:data}, $N_p=2$ is the number of polarization channels, and $W_i$ is the intrinsic burst width, which we derive from the initial detection and list in table~\ref{table:limit}.
$W_b$ is the broadened burst width, calculated using:
\begin{equation}
    W_b=\sqrt{W_i^{2}+t_{\rm samp}^{2}+t_{\rm chan}^{2}+t_{\rm scatt}^{2}},
	\label{eq:wb}
\end{equation}
where $t_{\rm samp}$ is the sampling time, $t_{\rm scatt}$ is the scattering
time~\footnote{We set $t_{\rm scatt} = 0$ because the S/N of the initial detected bursts were too low to allow a reliable measurement of the scattering time.}, and $t_{\rm chan}$ is the per-channel dispersive delay (intra-channel DM smearing), 
which can be estimated as~\citep{dmsmearing}:
\begin{equation}
    t_{\rm chan}=8.3{\rm \mu s}\times\left(\frac{\Delta\nu_{\rm chan}}{\rm MHz}\right)\times\left(\frac{\nu_c}{\rm GHz}\right)^{-3}\times\left(\frac{\rm DM}{{\rm pc}\cdot{\rm cm}^{-3}}\right),
	\label{eq:dmsmear}
\end{equation}
where $\Delta\nu_{\rm chan}$ is the frequency channel width in MHz and $\nu_c$ is the center observing frequency in GHz.

\subsection{Poisson Distribution} \label{Dis_Po}
If the bursts from a source are Poisson-distributed with event rate $r$, the probability of detecting N photons in the observation time T is:
\begin{equation}
    P(N|rT)=\frac{(rT)^{N}e^{-rT}}{N!},
	\label{eq:possion}
\end{equation}

To obtain the repetition rate, we follow the scaling process presented by \cite{fast36repeat} and derive the Poisson-scaled repeating rate $r_{\rm scaled}$. This scaled rate is based on the assumption that FRBs occur randomly in time. It includes a sensitivity scaling factor to enable comparison between different observations:
\begin{equation}
    r_{\rm scaled}=\frac{N_{\rm bursts}}{T_m(S_i/S_0)^{-1.5}+P_{\rm acc}T_f(S_f/S_0)^{-1.5}},
	\label{eq:rscale}
\end{equation}
where $N_{\rm bursts}$ is the number of detected bursts, $T_m$ is the on-source exposure time in hours, and $S_i$ is the sensitivity limit in Jansky of previous observations~(the Parkes multibeam observation from 1997 to 2001). $T_f$ and $S_f$ are the on-source exposure time and sensitivity limit of our follow up observations with Parkes and FAST. $S_0=1~\rm Jy$ is the reference flux density, and the power-law index $\alpha=-1.5$ is the $N \propto S^{\alpha}_{min}$ relation. The $P_{\rm acc}=0.25$ is the probability of the source being within the field of view for FAST Snapshot mode observations.%~\citep{fast36repeat}

\subsection{Weibull Distribution}\label{Dis_Wei}
If the bursts from a source are Weibull-distributed, its probability density function of is
\begin{equation}
   \mathcal{W}(\delta|k,r)=\frac{k}{\delta}\left[r\delta\Gamma\left(1+\frac{1}{k}\right)\right]^{k}e^{-\left[r\delta\Gamma\left(1+\frac{1}{k}\right)\right]^{k}},\label{eq:weibull}
\end{equation}
where $\delta$ is the interval between bursts, and the $\Gamma$ function is defined as $\Gamma(x) = \int_0^\infty t^{x-1} e^{-t} \, \mathrm{d}t$. Here, $k$ is the shape parameter and $r$ is the burst rate. 

%Taking a prior for $k$ and $r$ into account, the posterior distribution for the two parameters is
%\begin{equation}
%\mathcal{P}(k, r \mid N, t_1, \ldots, t_N) \propto \mathcal{P}(N, t_1, \ldots, t_N \mid k, r) \mathcal{P}(k, r),
%\end{equation}
%where 
%\begin{equation}
%    \mathcal{P}(k, r) \propto k^{-1} r^{-1}\label{eq:weibullprior}
%\end{equation}
%is the Jeffrey’s prior, the $t_N$ is the arrival time of number $N$ burst, and the log$k$ and log$r$ are uniform sampled~($10^{-2}<\text{log}k<10^{1}$,$10^{-5}<\text{log}r<10^{1}$).

%Following the formula from~\cite{weibull2018}, the probability density for observing N bursts in a single observation is
%\begin{equation}
%\begin{aligned}
%\mathcal{P}(N, t_1, \ldots, t_N | k, r) = {} & r \cdot \text{CDF}(t_1 | k, r) \\
%&\times \text{CDF}(\Delta - t_N | k, r) \\
%& \times \prod_{i=1}^{N-1} \mathcal{W}(t_{i+1} - t_i | k, r),\label{eq:weibullNN}
%\end{aligned}
%\end{equation}
%where $\text{CDF}(\delta \mid k, r)=\text{exp}\left({-\left[r\delta\Gamma\left(1+\frac{1}{k}\right)\right]^{k}}\right)$, $\Delta$ is length of an observation.

Following the formula from~\cite{weibull2018}, the probability density for observing $N=1$ bursts in a single observation is
%The probability density for observing $N=1$ bursts in a single observation is
\begin{equation}
\begin{aligned}
\mathcal{P}(N=1 \mid k, r) = {} & r \cdot \text{CDF}(t_{\mathrm{i}} \mid k, r) \\
&\times \text{CDF}(\Delta - t_{N} \mid k, r),\label{eq:weibullN1}
\end{aligned}
\end{equation}
where $\text{CDF}(\delta \mid k, r)=\text{exp}\left({-\left[r\delta\Gamma\left(1+\frac{1}{k}\right)\right]^{k}}\right)$, $\Delta$ is length of an observation.

The probability density for observing $N=0$ bursts in a single observation is
\begin{equation}
P(N = 0 \mid k, r) = \frac{\Gamma_i\left(1/k, (\Delta \, r \, \Gamma(1 + 1/k))^k\right)}{k \, \Gamma(1 + 1/k)},\label{eq:weibullN0}
\end{equation}
%where 
%\begin{equation}
%\Gamma_i(x,z) = \int_z^\infty \mathrm{d}t \, t^{x-1} \mathrm{e}^{-t}.\label{eq:weibullgammai}
%\end{equation}
%\\
The observation length $\Delta$ is scaled by the sensitivity and $P_{\rm acc}$, following method from~\cite{fast36repeat}:
\begin{equation}
\Delta_{\text{scaled}} = \Delta \cdot (S/S_0)^{-1.5} \cdot P_{\text{acc}},\label{eq:weibulldelta}
\end{equation}
where $S$ is the sensitivity limit derived from equation~(\ref{eq:slim}).

Using the equation~(\ref{eq:weibullN1}), equation~(\ref{eq:weibullN0}) and equation~(\ref{eq:weibulldelta}), we carried out a Bayesian inference using the nested sampling software package MultiNest~\citep{multinest,frb180301a1}, the parameters $\log k$ and $\log r$ were uniformly sampled ($10^{-2} < \log k < 10^{1}$, $10^{-5} < \log r < 10^{1}$).

\section{Results and Discussion}
\label{sec:discuss}

\subsection{Limits of follow-up observation}
After processing all 65.3 hours of data (59.0 hours from Parkes UWL and 6.3 hours from FAST), no convincing repeating bursts were detected from any of the targeted sources.

The derived flux limits are listed in the second and third columns of table~\ref{table:limit}. Here, $S_i$ denotes the flux density of the initial detection,
while $S_f$ represents the sensitivity limit of our follow-up observations with Parkes and FAST. Sources observed by FAST are marked with an asterisk ($^*$) in the source name column.

\begin{figure*}
    \begin{center}
    \includegraphics[width=0.9\textwidth]{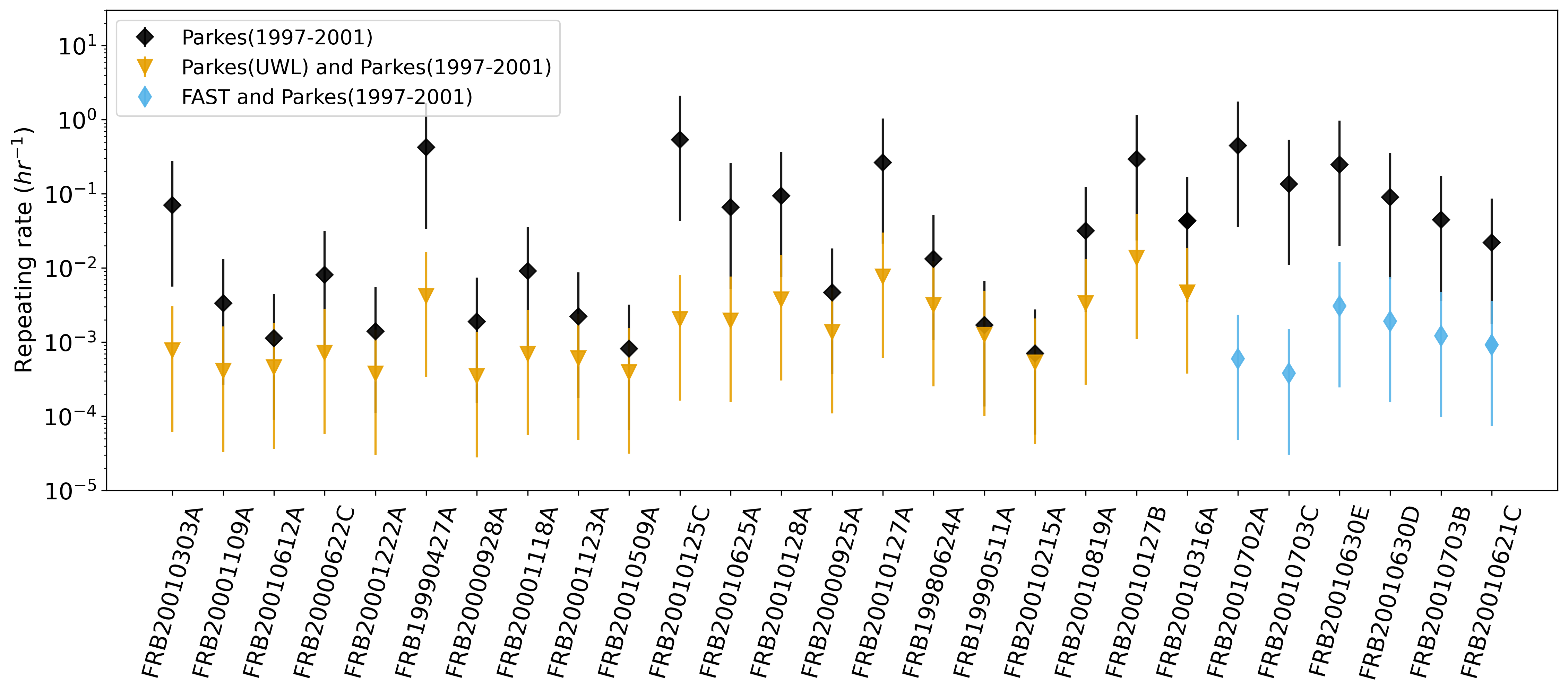}
    \caption{Limits of Poisson repeating rate $r_{\rm scaled}$ of 27 FRBs. The error bars represent 90\% confidence intervals, calculated using the method from~\cite{possionconfidence} to provide improved accuracy for low Poisson event rates.}
    \label{figure:possion}
    \end{center}
\end{figure*}

Considering the derived flux limit, the resulting upper limits on repetition rates above 1\,Jy under a Poisson process are shown in Figure~\ref{figure:possion}
~(The 1\,Jy was chosen as the reference flux for consistent with \citealp{good2023} and \citealp{fast36repeat}). The observations from the different surveys are combined to derive the rate using equation~\ref{eq:rscale}.
The rates are plotted on a logarithmic $y$-axis, with error bars representing 90\% confidence intervals. 
Data from the original Parkes observations (1997$-$2001) and their combinations with Parkes UWL and FAST are shown as blue, green, and red triangles, respectively.
The combined limits from both follow-up and prior archival observations span a range of approximately $10^{-3.5}$ to $10^{-1.9}\, h^{-1}$.

\begin{figure*}
    \begin{center}
    \includegraphics[width=0.9\textwidth]{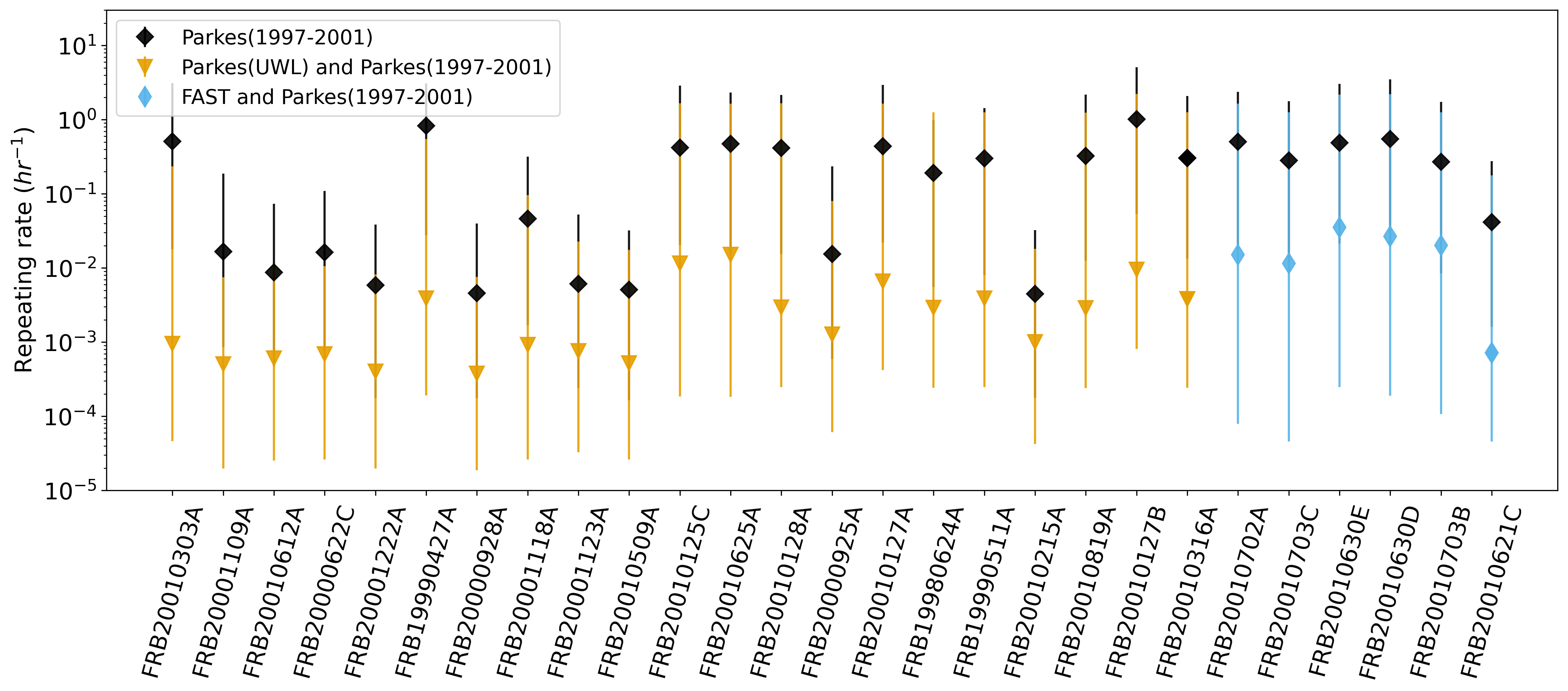}
    \caption{Limits of Weibull repeating rate of 27 FRBs. The error bars represent 90\% confidence intervals from the posterior probability distribution of $\text{log}\,r$.}
    \label{figure:weibull}
    \end{center}
\end{figure*}

In addition to the Poisson analysis, the upper limits on the repetition rates above 1\,Jy under a Weibull process (see calculation details in section~\ref{Dis_Wei}) are presented in Figure~\ref{figure:weibull}. The error bars indicate the 90\% confidence intervals from the posterior probability distribution of $\log r$. 
The combined limits from follow-up observations and prior archival observations range from approximately $10^{-3.4}$ to $10^{-1.5}\, h^{-1}$ under the Weibull process assumption.

Compared to 10-min per-source exposure time in the FAST campaign by~\cite{fast36repeat}, our observations achieved significantly longer integration times, up to 2 hours per source. In total, we conducted 6.3 hours of FAST observations covering six sources and 59 hours of Parkes observations targeting 21 sources.
The resulting repetition-rates limit of $10^{-3.5}$ to $10^{-1.9}\, h^{-1}$ is approximately an order of magnitude stricter than that obtained by~\cite{fast36repeat}, which was $10^{-2.6}$ to $10^{-0.22}\, h^{-1}$. 
%work conducted longer observations for each source which typically exceeding 40 minutes. We execute observations of 6.3 hours for 6 sources with FAST and a total of 59 hours for 21 sources using the Parkes telescope. The repetition rate upper limit for our observations is $\sim 10^{-1.9}$ that are an order of magnitude lower than \cite{fast36repeat}.

\begin{table*}[ht]
  \begin{scriptsize}
  \caption{Overview of sensitivity and repeating rate limits. Sources observed by FAST 19-beam receiver are marked with an asterisk ($^*$) in the source name column, the other sources are observed by Parkes UWL receiver. $S_i$ is the sensitivity limit in Jansky of previous observations~(the Parkes multibeam observation from 1997 to 2001). $S_f$ is the sensitivity limit of our follow up observations with Parkes or FAST. $r_{\rm scaled}$(P) is the rate limit under Poisson model, $r_{\rm scaled}$(W) is the rate limit under Weibull model.\label{table:limit}}
  \renewcommand\arraystretch{1.0}
  \setlength{\tabcolsep}{2.0mm}  
  \begin{center}
  \begin{tabular}{@{\extracolsep{\fill}}cccccccccc}
  \hline  
\multicolumn{1}{c}{}  &  &   &        &  Parkes(1997-2001)  & Parkes(1997-2001)   &   Parkes(1997-2001)   &  Follow-up  &  Follow-up   &  Follow-up  \\
\multicolumn{1}{c}{Source}  & $W_i$ & $S_i$ &  $S_f$   & $r_{\rm scaled}$(P)    & $r_{\rm scaled}$(W) &  k  & $r_{\rm scaled}$(P)    & $r_{\rm scaled}$(W) &  k  \\
 \multicolumn{1}{c}{Name}   & (ms) &(Jy) & (Jy)  &  $(\rm hr^{-1})$ & $(\rm hr^{-1})$ & & $(10^{-3}\rm hr^{-1})$ &$(10^{-3}\rm hr^{-1})$ & \\
\hline
FRB20010702A* & 5.1 & 0.214 & 0.004 & $0.447^{+1.310}_{-0.411}$ & $0.504^{+1.869}_{-0.486}$ & $1.0^{+8.5}_{-0.7}$ & $0.6^{+1.8}_{-0.6}$ & $15.1^{+1636.9}_{-15.0}$ & $0.2^{+2.2}_{-0.1}$ \\ 
FRB20010703C* & 9.2 & 0.154 & 0.003 & $0.136^{+0.400}_{-0.125}$ & $0.281^{+1.498}_{-0.272}$ & $0.9^{+8.7}_{-0.6}$ & $0.4^{+1.1}_{-0.3}$ & $11.5^{+1246.9}_{-11.5}$ & $0.2^{+1.8}_{-0.1}$ \\ 
FRB20010630E* & 2.9 & 0.342 & 0.006 & $0.247^{+0.724}_{-0.227}$ & $0.490^{+2.560}_{-0.469}$ & $5.9^{+3.6}_{-5.6}$ & $3.1^{+9.0}_{-2.8}$ & $35.2^{+2156.4}_{-35.0}$ & $0.2^{+7.4}_{-0.2}$ \\ 
FRB20010630D* & 5.4 & 0.195 & 0.004 & $0.090^{+0.264}_{-0.083}$ & $0.551^{+2.953}_{-0.529}$ & $2.6^{+7.0}_{-2.2}$ & $1.9^{+5.6}_{-1.8}$ & $26.8^{+2168.1}_{-26.6}$ & $0.2^{+7.5}_{-0.1}$ \\ 
FRB20010703B* & 9.8 & 0.141 & 0.003 & $0.045^{+0.132}_{-0.041}$ & $0.270^{+1.477}_{-0.262}$ & $2.4^{+7.1}_{-2.2}$ & $1.2^{+3.6}_{-1.1}$ & $20.1^{+1238.6}_{-20.0}$ & $0.2^{+5.4}_{-0.1}$ \\ 
FRB20010621C* & 13.9 & 0.139 & 0.003 & $0.022^{+0.065}_{-0.020}$ & $0.041^{+0.233}_{-0.040}$ & $3.1^{+6.4}_{-2.8}$ & $0.9^{+2.7}_{-0.8}$ & $0.7^{+176.5}_{-0.7}$ & $0.3^{+7.5}_{-0.2}$ \\ 
FRB20010303A & 7.2  & 0.166 & 0.031 & $0.071^{+0.207}_{-0.065}$ & $0.513^{+2.606}_{-0.495}$ & $0.8^{+8.7}_{-0.5}$ & $0.8^{+2.3}_{-0.7}$ & $1.0^{+234.1}_{-0.9}$ & $0.2^{+9.3}_{-0.1}$ \\ 
FRB20001109A & 17.0  & 0.142 & 0.020 & $0.003^{+0.010}_{-0.003}$ & $0.017^{+0.171}_{-0.016}$ & $2.1^{+7.4}_{-1.8}$ & $0.4^{+1.2}_{-0.4}$ & $0.5^{+7.0}_{-0.5}$ & $0.4^{+9.1}_{-0.2}$ \\ 
FRB20010612A & 10.5  & 0.142 & 0.025 & $0.001^{+0.003}_{-0.001}$ & $0.009^{+0.064}_{-0.008}$ & $1.3^{+8.2}_{-1.0}$ & $0.5^{+1.3}_{-0.4}$ & $0.6^{+10.6}_{-0.6}$ & $0.4^{+9.1}_{-0.2}$ \\ 
FRB20000622C & 10.5  & 0.141 & 0.025 & $0.008^{+0.024}_{-0.007}$ & $0.016^{+0.093}_{-0.016}$ & $3.0^{+6.5}_{-2.6}$ & $0.7^{+2.1}_{-0.7}$ & $0.7^{+9.9}_{-0.7}$ & $0.4^{+9.1}_{-0.2}$ \\ 
FRB20001222A & 20.4  & 0.120 & 0.018 & $0.001^{+0.004}_{-0.001}$ & $0.006^{+0.033}_{-0.006}$ & $0.8^{+8.8}_{-0.5}$ & $0.4^{+1.1}_{-0.3}$ & $0.4^{+7.8}_{-0.4}$ & $0.5^{+9.0}_{-0.3}$ \\ 
FRB19990427A & 1.3  & 1.232 & 0.072 & $0.424^{+1.243}_{-0.390}$ & $0.832^{+2.243}_{-0.804}$ & $5.0^{+4.5}_{-4.6}$ & $4.2^{+12.3}_{-3.9}$ & $3.9^{+544.8}_{-3.7}$ & $0.3^{+9.2}_{-0.2}$ \\ 
FRB20000928A & 39.5  & 0.087 & 0.013 & $0.002^{+0.006}_{-0.002}$ & $0.005^{+0.035}_{-0.004}$ & $5.9^{+3.6}_{-5.6}$ & $0.3^{+1.0}_{-0.3}$ & $0.4^{+7.2}_{-0.4}$ & $0.4^{+9.2}_{-0.2}$ \\ 
FRB20001118A & 18.7  & 0.108 & 0.019 & $0.009^{+0.027}_{-0.008}$ & $0.046^{+0.273}_{-0.045}$ & $9.5^{+0.0}_{-9.2}$ & $0.7^{+2.0}_{-0.6}$ & $0.9^{+95.8}_{-0.9}$ & $0.2^{+8.3}_{-0.1}$ \\ 
FRB20001123A & 18.1  & 0.137 & 0.019 & $0.002^{+0.007}_{-0.002}$ & $0.006^{+0.047}_{-0.006}$ & $1.3^{+8.3}_{-1.0}$ & $0.6^{+1.8}_{-0.6}$ & $0.8^{+22.0}_{-0.7}$ & $0.3^{+9.2}_{-0.1}$ \\ 
FRB20010509A & 26.1  & 0.107 & 0.016 & $0.001^{+0.002}_{-0.001}$ & $0.005^{+0.027}_{-0.005}$ & $2.6^{+6.9}_{-2.3}$ & $0.4^{+1.2}_{-0.4}$ & $0.5^{+17.1}_{-0.5}$ & $0.4^{+8.2}_{-0.2}$ \\ 
FRB20010125C & 7.7  & 0.242 & 0.030 & $0.539^{+1.579}_{-0.496}$ & $0.422^{+2.464}_{-0.401}$ & $4.7^{+4.8}_{-4.4}$ & $2.0^{+6.0}_{-1.9}$ & $11.6^{+1647.5}_{-11.4}$ & $0.2^{+6.8}_{-0.1}$ \\ 
FRB20010625A & 8.7  & 0.197 & 0.028 & $0.066^{+0.193}_{-0.061}$ & $0.476^{+1.844}_{-0.460}$ & $7.1^{+2.4}_{-6.8}$ & $2.0^{+5.7}_{-1.8}$ & $15.1^{+1638.7}_{-15.0}$ & $0.2^{+6.1}_{-0.1}$ \\ 
FRB20010128A & 6.3  & 0.180 & 0.033 & $0.094^{+0.276}_{-0.087}$ & $0.417^{+1.749}_{-0.402}$ & $1.2^{+8.3}_{-0.9}$ & $3.8^{+11.1}_{-3.5}$ & $2.9^{+1654.7}_{-2.7}$ & $0.3^{+7.5}_{-0.1}$ \\ 
FRB20000925A & 20.0  & 0.120 & 0.018 & $0.005^{+0.014}_{-0.004}$ & $0.016^{+0.220}_{-0.015}$ & $6.5^{+3.1}_{-6.1}$ & $1.4^{+4.0}_{-1.3}$ & $1.3^{+78.7}_{-1.2}$ & $0.4^{+8.2}_{-0.2}$ \\ 
FRB20010127A & 3.4  & 0.240 & 0.044 & $0.265^{+0.777}_{-0.244}$ & $0.439^{+2.501}_{-0.417}$ & $1.7^{+7.8}_{-1.4}$ & $7.7^{+22.4}_{-7.0}$ & $6.6^{+1637.0}_{-6.2}$ & $0.3^{+7.4}_{-0.2}$ \\ 
FRB19980624A & 10.1  & 0.146 & 0.026 & $0.013^{+0.039}_{-0.012}$ & $0.191^{+0.788}_{-0.186}$ & $7.1^{+2.4}_{-6.8}$ & $3.2^{+9.3}_{-2.9}$ & $2.9^{+1257.6}_{-2.7}$ & $0.2^{+6.8}_{-0.1}$ \\ 
FRB19990511A & 9.3  & 0.187 & 0.027 & $0.002^{+0.005}_{-0.002}$ & $0.301^{+1.126}_{-0.293}$ & $7.1^{+2.4}_{-6.8}$ & $1.3^{+3.7}_{-1.2}$ & $3.9^{+1259.0}_{-3.6}$ & $0.2^{+9.3}_{-0.1}$ \\ 
FRB20010215A & 38.1  & 0.085 & 0.013 & $0.001^{+0.002}_{-0.001}$ & $0.004^{+0.028}_{-0.004}$ & $5.3^{+4.3}_{-5.0}$ & $0.5^{+1.6}_{-0.5}$ & $1.0^{+17.2}_{-1.0}$ & $0.5^{+9.0}_{-0.3}$ \\ 
FRB20010819A & 18.3  & 0.130 & 0.019 & $0.032^{+0.093}_{-0.029}$ & $0.326^{+1.855}_{-0.314}$ & $4.9^{+4.6}_{-4.6}$ & $3.4^{+9.8}_{-3.1}$ & $2.9^{+1246.6}_{-2.6}$ & $0.2^{+6.1}_{-0.1}$ \\ 
FRB20010127B & 3.1  & 0.311 & 0.047 & $0.294^{+0.860}_{-0.270}$ & $1.014^{+4.062}_{-0.961}$ & $1.7^{+7.9}_{-1.3}$ & $13.7^{+40.2}_{-12.6}$ & $9.5^{+2214.9}_{-8.7}$ & $0.3^{+9.2}_{-0.1}$ \\ 
FRB20010316A & 13.0  & 0.132 & 0.023 & $0.043^{+0.127}_{-0.040}$ & $0.305^{+1.782}_{-0.291}$ & $1.8^{+7.8}_{-1.5}$ & $4.7^{+13.8}_{-4.3}$ & $3.8^{+1251.8}_{-3.6}$ & $0.2^{+9.3}_{-0.1}$ \\ 
\hline
\end{tabular}
\end{center}
\end{scriptsize}
\end{table*}

\subsection{Comparison with bright FRBs discovered by Parkes}

%
%Based on the observation parameters and follow-up campaigns for the bright, apparently non-repeating FRBs (S/N $\ge 10$) discovered by Parkes, as summarized in~\cite{songborrat}, along with the long-term monitoring of the repeating FRB 20180301A reported by~\cite{frb180301a1} and~~\cite{frb20180301a2}, we analyzed the follow-up observations for 27 Parkes bright FRBs and derived their repetition rate limits~\footnote{For three of the 30 bright Parkes FRBs, no archival datasets are available to constrain their rate limits.} under a Poisson distribution, as shown in Figure~\ref{figure:possion27frb}.

\begin{figure*}
    \begin{center}
    \includegraphics[width=0.9\textwidth]{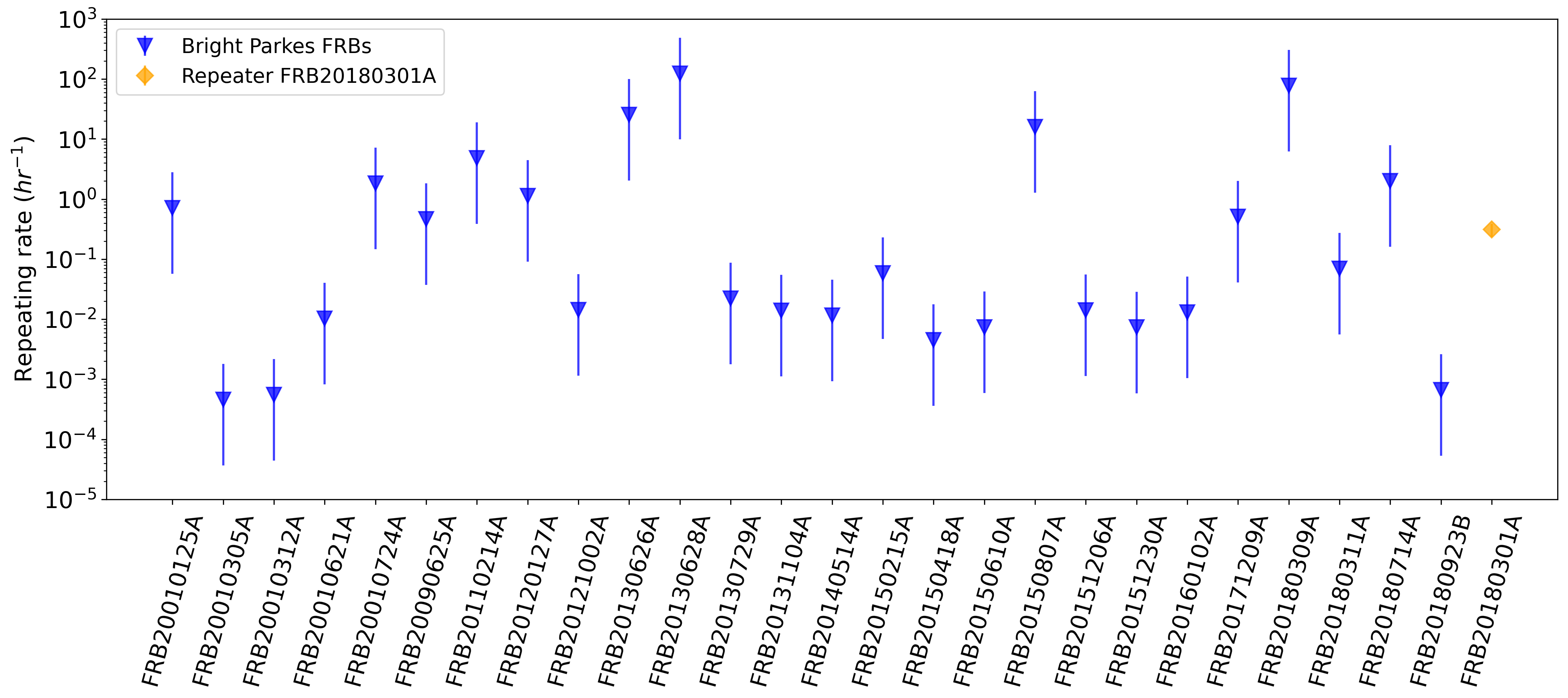}
    \caption{Limits of Poisson repeating rate $r_{\rm scaled}$ of 27 published FRBs from Parkes~(26 apparently non-repeating FRBs and the repeating FRB~20180301A). The error bars represent 90\% confidence intervals, calculated using the method from~\cite{possionconfidence} to provide improved accuracy for low Poisson event rates.}
    \label{figure:possion27frb}
    \end{center}
\end{figure*}

Our comparison is based on the observation parameters and follow-up campaigns for bright, apparently non-repeating FRBs (S/N $\ge 10$) discovered by the Parkes telescope.
A summary of these FRBs is provided by ~\cite{songborrat}. 
For example, observations of FRB 010724A with the Parkes multibeam receiver comprise of three kinds of observation setups: 42.0 hours at 1526 MHz with 512 MHz bandwidth, 512 frequency channels, and 96 $\rm \mu s$ sampling; 7.3 hours at 1374 MHz with 288 MHz bandwidth, 96 channels, and 1000 $\rm \mu s$ sampling; and 3.6 hours at 1382 MHz with 400 MHz bandwidth, 1024 frequency channels, and 64 $\rm \mu s$ sampling~\footnote{More detailed observation informations for the FRBs can be found in \url{https://data.csiro.au/search/keyword}.}.
We also incorporated long-term monitoring of the repeating FRB 20180301A, as reported by ~\cite{frb180301a1} and ~\cite{frb20180301a2}.
Using these information, we analyzed the follow-up observations for 27 bright Parkes FRBs.
Under a Poisson distribution, we derived repetition rate limits for these events~\footnote{For three of the 30 bright Parkes FRBs, no archival datasets are available to constrain their rate limits.}.
The results are shown in Figure~\ref{figure:possion27frb}.
In contrast to these previous heterogeneous datasets, 

our observations were conducted with consistently configured instrumental parameters and comparable integration times, thereby obtaining a new dataset that more reliably characterizes the true repetition behavior of these FRBs.

\begin{figure*}
    \begin{center}
    \includegraphics[width=0.9\textwidth]{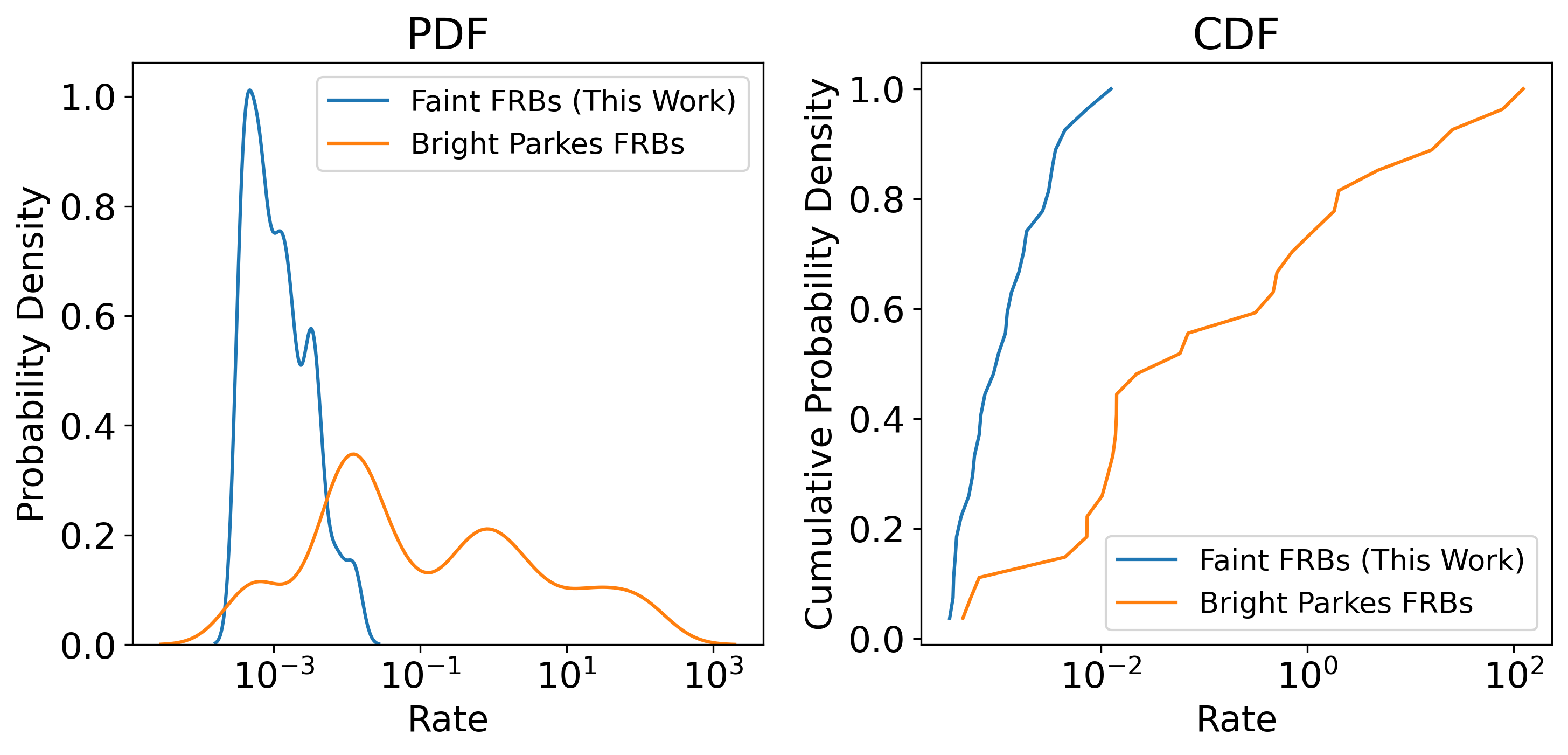}
    \caption{The PDF and CDF comparssion of the rate limit from this work and Parkes previous work.}
    \label{figure:pdfcdf}
    \end{center}
\end{figure*}

Figure~\ref{figure:pdfcdf} shows the Probability Density Function~(PDF) and Cumulative Distribution Function~(CDF) of the repetition rate limit for the faint FRBs from the PTD-I and the bright Parkes FRBs. The PDF is estimated by Kernel Density Estimation~(KDE) method. 
As shown in Figure 4, our systematically derived repetition rate constraints span a relatively narrow range of
$\sim 10^{-3.5}$ to $10^{-1.9} h^{-1}$
under the Poisson model, which is approximately an order of magnitude tighter than the upper limits obtained for the bright Parkes FRBs ($\sim 10^{-4}$ to $10^{-2} h^{-1}$).
This improvement in stringent constraints results from the longer integration times (up to 2 hours per source) and the enhanced sensitivity provided by FAST and the Parkes UWL receiver. 
%These substantially stricter limits allow us to more confidently rule out moderate repetition activity in these sources, providing stronger evidence for the existence of a population of genuinely non-repeating FRBs or sources with extremely low intrinsic activity rates.

These substantially stricter limits allow us to more confidently rule out moderate repetition in these sources. This provides stronger evidence for the existence of a population of genuinely non-repeating FRBs. It also suggests that some sources may have extremely low intrinsic activity rates.
Moreover, our constraints have converged into a well-defined range. This convergence was made possible by applying consistent observational and analytical frameworks across all sources. It suggests that the faint Parkes FRBs form a relatively homogeneous population with similar repetition characteristics, unlike the broad dispersion seen in bright Parkes FRBs studies that used heterogeneous methodologies.

%Moreover, the convergence of our constraints into a well-defined range, which is made possible by applying consistent observational and analytical frameworks across all sources, suggests that the faint Parkes FRBs constitute a relatively homogeneous population with similar repetition characteristics, rather than exhibiting the broad dispersion seen in earlier studies that employed heterogeneous methodologies.

Understanding FRB populations requires accounting for their diverse burst rates.
The Galactic event FRB 200428 from SGR 1935+2154 is illustrative: its energy was $\sim$30 times weaker than the faintest extragalactic FRBs~\citep{frb200428, STARE2}, and would appear as a non-repeater at extragalactic distances.
%When considered at extragalactic distances, such events would fall below detection thresholds, appearing as non-repeaters. 
This supports the inference by \cite{volume_rate} that most FRBs are repeaters given their high volumetric rate. Our stringent rate limits further agree with \cite{121102_strange} and indicate that hyperactive repeaters such as FRB~20121102A and FRB~20201124A represent a distinct population from apparently non-repeating FRBs.

\section{Conclusions}
\label{sec:Conclusions}
Based on 65.3 hours of sensitive follow-up observations with FAST and the Parkes UWL receiver, we report no detection of repeating bursts from 27 FRBs identified in the PTD-I. From these non-detections, we derived strict upper limits on their repetition rates, ranging from $\sim 10^{-3.5}$ to $10^{-1.9} h^{-1}$ under a Poisson model, and $\sim 10^{-3.4}$ to $10^{-1.5} h^{-1}$ under a Weibull model.

The combination of the Parkes UWL and FAST 19-beam observations provides improved sensitivity and bandwidth coverage. Given FAST’s exceptional sensitivity, it could detect bursts more than 20 times fainter than Parkes. The continued non-detections strongly suggest that these sources are either non-repeating on human timescales or have repetition rates below $\sim 10^{-3.5} h^{-1}$. These may represent ``inactive repeaters'' with extremely low intrinsic activity.

By employing consistent observational setups, data processing, and statistical analyses, we obtained uniform and comparable results across all sources. This consistency underscores the importance of methodological uniformity in FRB follow-up studies, which helps minimize systematic biases and enables a more reliable assessment of the underlying FRB population.
Nonetheless, some limitations remain. The total on-source time per FRB is modest, and possible temporal clustering of burst activity means active phases may have been missed. In addition, as some repeaters show narrow-band emission~\citep{narrow,narrow2}, bursts outside the FAST 19-beam receiver’s frequency coverage cannot be excluded.

Future long-term monitoring campaigns with highly sensitive telescopes such as FAST will be essential to further tighten these rate constraints and potentially capture rare bursts from these apparently quiet sources. Confirming or definitively ruling out repetition in such faint FRBs will be a crucial step toward understanding their progenitors and the true nature of the FRB population.

\section*{DATA AVAILABILITY STATEMENTS}
The dataset P1040 and P1105 used in this work is available from the CSIRO Data Access Portal, \url{https://data.csiro.au/}. P1040:
\dataset[doi.org/10.25919/ngms-ph41]{https://doi.org/10.25919/ngms-ph41}, P1105 \dataset[doi.org/10.25919/mefz-c228]{https://doi.org/10.25919/mefz-c228}.  Detailed parameters for the FRB sources from PTD-I are available at~\url{https://astroyx.github.io/}.

\section*{Acknowledgments}
This work is partially supported by the National Natural Science Foundation of China (grant Nos. 12321003, 12041306, 12273113, 12233002, 12003028), the international Partnership Program of Chinese Academy of Sciences for Grand Challenges (114332KYSB20210018), the National SKA Program of China (2022SKA0130100), the National Key R\&D Program of China (2021YFA0718500), Postdoctoral Fellowship Program of CPSF (grant No. GZC20252100), the ACAMAR Postdoctoral Fellow, China Postdoctoral Science Foundation (grant No. 2025M773201), and Jiangsu Funding Program for Excellent Postdoctoral Talent. \\


\bibliography{main}
\bibliographystyle{aasjournal}

\appendix
%\section{APPENDIX}
\label{appendix}

\section{Observation epoch}

\begin{table*}[ht]
\renewcommand{\thetable}{A.1} % 自定义编号
  \begin{scriptsize}
  \caption{The observation date and telescope of 27 FRBs. \label{table:obslog}}
  \renewcommand\arraystretch{0.95}
  \setlength{\tabcolsep}{2.0mm}  
  \begin{center}
  \begin{tabular}{@{\extracolsep{\fill}}ccccccccc}
  \hline  
\multicolumn{1}{c}{Source}  & Telescope  &    Start Date(UTC)     &  Integration  Time(h)         \\
\hline
\multirow{1}{*}{FRB20010702A} & FAST & 2021-08-09 12:57:29.267 & 2.0 \\
\hline
\multirow{1}{*}{FRB20010703C} & FAST & 2021-08-13 13:29:00.000 & 2.0 \\
\hline
\multirow{1}{*}{FRB20010630E} & FAST & 2025-08-24 12:24:00.000 & 0.58 \\
\hline
\multirow{1}{*}{FRB20010630D} & FAST & 2025-08-24 11:37:00.000 & 0.58 \\
\hline
\multirow{1}{*}{FRB20010703B} & FAST & 2025-08-24 10:50:00.000 & 0.58 \\
\hline
\multirow{1}{*}{FRB20010621C} & FAST & 2025-09-05 13:58:00.000 & 0.58 \\
\hline
\multirow{6}{*}{FRB20010303A} & Parkes & 2021-01-02 11:56:43.101 & 1.7 \\
\cline{2-4}
            & Parkes & 2021-01-16 08:34:23.101 & 1.01\\
\cline{2-4}
            & Parkes & 2021-01-24 06:59:06.101 & 1.51\\
\cline{2-4}
            & Parkes & 2021-01-25 11:07:54.101 & 0.85\\
\cline{2-4}
            & Parkes & 2021-01-26 08:23:53.101 & 0.57\\
\cline{2-4}
            & Parkes & 2021-04-04 01:22:29.000 & 1.23\\
\hline
\multirow{6}{*}{FRB20001109A} & Parkes & 2021-01-03 08:13:33.103 & 1.01 \\
\cline{2-4}
            & Parkes & 2021-01-21 06:50:13.105 & 1.16\\
\cline{2-4}
            & Parkes & 2021-01-26 06:48:53.103 & 1.51\\
\cline{2-4}
            & Parkes & 2021-01-27 07:06:24.103 & 1.01\\
\cline{2-4}
            & Parkes & 2021-02-27 05:48:45.103 & 1.01\\
\cline{2-4}
            & Parkes & 2022-08-29 14:49:06.113 & 0.34\\
\hline
\multirow{4}{*}{FRB20010612A} & Parkes & 2020-12-29 09:15:45.103 & 1.01 \\
\cline{2-4}
            & Parkes & 2021-01-02 09:45:06.103 & 1.72\\
\cline{2-4}
            & Parkes & 2021-01-16 10:41:56.103 & 1.01\\
\cline{2-4}
            & Parkes & 2021-01-21 05:17:36.103 & 1.51\\
\hline
\multirow{4}{*}{FRB20000622C} & Parkes & 2021-01-02 07:22:32.102 & 1.59 \\
\cline{2-4}
            & Parkes & 2021-01-03 07:09:11.102 & 1.02\\
\cline{2-4}
            & Parkes & 2021-01-16 09:39:12.102 & 1.01\\
\cline{2-4}
            & Parkes & 2021-01-26 05:16:12.102 & 1.51\\
\hline
\multirow{5}{*}{FRB20001222A} & Parkes & 2020-12-29 11:22:06.103 & 0.87 \\
\cline{2-4}
            & Parkes & 2021-01-03 10:19:47.103 & 1.62\\
\cline{2-4}
            & Parkes & 2021-01-24 08:33:17.103 & 0.88\\
\cline{2-4}
            & Parkes & 2021-01-27 08:10:16.103 & 0.82\\
\cline{2-4}
            & Parkes & 2021-02-27 06:52:39.103 & 0.59\\
\hline
\multirow{3}{*}{FRB19990427A} & Parkes & 2020-12-26 16:14:12.105 & 1.75 \\
\cline{2-4}
            & Parkes & 2021-04-04 09:19:04.105 & 1.01\\
\cline{2-4}
            & Parkes & 2021-09-18 00:13:36.105 & 1.77\\
\hline
\multirow{3}{*}{FRB20000928A} & Parkes & 2020-12-29 10:19:18.105 & 1.01 \\
\cline{2-4}
            & Parkes & 2021-01-03 09:17:07.105 & 1.01\\
\cline{2-4}
            & Parkes & 2021-01-24 05:12:38.105 & 1.51\\
\hline
\multirow{2}{*}{FRB20001118A} & Parkes & 2021-01-03 03:08:57.104 & 1.81 \\
\cline{2-4}
            & Parkes & 2021-09-19 03:17:16.104 & 1.69\\
\hline
\multirow{2}{*}{FRB20001123A} & Parkes & 2021-04-04 02:44:38.102 & 1.01 \\
\cline{2-4}
            & Parkes & 2022-03-12 08:41:15.102 & 2.24\\
\hline
\multirow{2}{*}{FRB20010509A} & Parkes & 2021-09-17 19:04:52.104 & 2.5 \\
\cline{2-4}
            & Parkes & 2025-01-23 05:20:35.104 & 0.22\\
\hline
\multirow{1}{*}{FRB20010125C} & Parkes & 2021-09-17 21:40:12.104 & 2.5 \\
\hline
\multirow{1}{*}{FRB20010625A} & Parkes & 2021-09-19 19:55:27.103 & 2.31 \\
\hline
\multirow{1}{*}{FRB20010128A} & Parkes & 2021-04-06 07:06:28.101 & 1.51 \\
\hline
\multirow{1}{*}{FRB20000925A} & Parkes & 2021-04-06 08:40:10.101 & 1.29 \\
\hline
\multirow{1}{*}{FRB20010127A} & Parkes & 2021-04-07 08:16:55.101 & 1.19 \\
\hline
\multirow{1}{*}{FRB19980624A} & Parkes & 2021-04-07 07:09:21.101 & 1.0 \\
\hline
\multirow{1}{*}{FRB19990511A} & Parkes & 2020-12-30 22:04:06.102 & 0.92 \\
\hline
\multirow{1}{*}{FRB20010215A} & Parkes & 2021-01-16 11:45:28.107 & 0.72 \\
\hline
\multirow{1}{*}{FRB20010819A} & Parkes & 2021-09-19 19:07:03.101 & 0.71 \\
\hline
\multirow{1}{*}{FRB20010127B} & Parkes & 2020-12-30 21:06:08.101 & 0.7 \\
\hline
\multirow{1}{*}{FRB20010316A} & Parkes & 2021-04-04 03:49:11.103 & 0.66 \\
\hline
\end{tabular}
\end{center}
\end{scriptsize}
\end{table*}

\clearpage

\end{document}